\documentclass[universe,editorial,accept,pdftex,moreauthors]{Definitions/mdpi} 

\firstpage{1} 
\pubvolume{10}
\issuenum{4}
\articlenumber{184}
\pubyear{2024}
\copyrightyear{2024}
\datereceived{3 April 2024} 
\dateaccepted{8 April 2024} 
\datepublished{18 April 2024} 
\hreflink{https://doi.org/10.3390/\linebreak universe10040184} 

\Title{Special Issue on Modified Gravity Approaches to the Tensions of $\Lambda$CDM: Goals and Highlights}

\TitleCitation{Special Issue on Modified Gravity Approaches to the Tensions of $\Lambda$CDM: Goals and Highlights}

\Author{{Eleonora Di Valentino} {$^{1}$}\orcidA{}, Leandros Perivolaropoulos {$^{2,}$}{*}\orcidB{} and Jackson Levi Said {$^{3}$}\orcidC{}}

\AuthorNames{Eleonora Di Valentino, Leandros Perivolaropoulos and Jackson Levi Said}

\AuthorCitation{{Di Valentino}, E.; Perivolaropoulos, L.; Said, J.L.}

\address{%
$^{1}$ \quad School of Mathematics and Statistics, University of Sheffield, {Sheffield S10 2AH}, UK; 
 e.divalentino@sheffield.ac.uk\\
$^{2}$ \quad Department of Physics, University of Ioannina, {45110 Ioannina}, Greece\\
$^{3}$ \quad Institute of Space Sciences and Astronomy, University of Malta, {2080 Msida}, 
Malta; jackson.said@um.edu.mt}

\corres{Correspondence: leandros@uoi.gr}

\abstract{The Special Issue on \href{https://www.mdpi.com/journal/universe/special_issues/Gravit}{"Modified Gravity Approaches to the Tensions of $\Lambda$CDM"} in the Universe journal tackles significant challenges faced by the $\Lambda$CDM model, including discrepancies in the Hubble constant, growth rate of structures, and cosmological anisotropies. These issues suggest foundational cracks in the model, raising questions about the validity of General Relativity, dark energy, and cosmological principles at large scales. This collection brings together leading researchers to delve into Modified Gravity theories as potential solutions. Covering approaches from Scalar-Tensor theories to $f(R,T)$ gravity and beyond, each contribution presents innovative research aimed at addressing the limitations of the $\Lambda$CDM model. This Special Issue not only highlights the theoretical and empirical strengths of Modified Gravity models but also opens avenues for future investigations, emphasizing the synergy between theoretical advancements and observational evidence to deepen our cosmological understanding.}

\keyword{cosmology; $\Lambda$CDM model; modified gravity; Hubble tension; dark energy}

\begin{document}

\setcounter{section}{0} 
\section{Introduction}
\label{sec:introduction}

The standard cosmological model, known as $\Lambda$CDM, has been remarkably successful in providing a coherent and predictive framework for understanding the Universe's evolution, its large-scale structure, and cosmic microwave background (CMB) radiation~\cite{Planck:2018vyg,ACT:2023dou,SPT-3G:2022hvq,eBOSS:2020yzd,KiDS:2020suj,DES:2021wwk}. Central to this model are the cosmological constant $\Lambda$, representing dark energy responsible for the accelerated expansion of the Universe, and cold dark matter (CDM), which accounts for the gravitational scaffolding underlying galaxy formation and evolution. Despite its triumphs, the $\Lambda$CDM model is not without its challenges~\cite{Perivolaropoulos:2021jda,Abdalla:2022yfr}. In recent years, high-precision cosmological observations have revealed a series of tensions that question the completeness of the $\Lambda$CDM paradigm. These tensions---most notably, the discrepancy in the Hubble constant ($H_0$) measurements from local and early Universe observations~\cite{Verde:2019ivm,DiValentino:2020zio,DiValentino:2021izs,Vagnozzi:2023nrq,Dainotti:2023yrk,Akarsu:2024qiq,Kamionkowski:2022pkx,Dainotti:2021pqg,Perivolaropoulos:2021jda,Jedamzik:2020zmd,Vagnozzi:2021tjv}, the growth rate of cosmic structures~\cite{DiValentino:2020vvd,Philcox:2021kcw,Heisenberg:2022gqk,Benisty:2020kdt,Kilo-DegreeSurvey:2023gfr,KiDS:2020ghu,Joudaki:2019pmv,Troster:2019ean}, and the scale of cosmological \mbox{anisotropies~\cite{Aluri:2022hzs,Secrest:2020has,Uzan:2008qp,McConville:2023xav,Watkins:2023rll,Cowell:2022ehf,Dam:2022wwh,Krishnan:2022qbv,Bengaly:2024ree,Ebrahimian:2023svi,Perivolaropoulos:2023tdt}}---suggest potential shortcomings in our understanding of the fundamental components and laws governing the Universe.

The emergence of these tensions has spurred a vibrant discourse within the cosmological community, giving rise to the motivation behind this Special Issue on ``Modified Gravity Approaches to the Tensions of $\Lambda$CDM''. These inconsistencies may herald the advent of a new standard cosmological model, one rooted in physics beyond the current paradigm. As we stand on the brink of potentially revolutionary discoveries, it becomes imperative to explore avenues beyond the conventional framework, questioning the validity of general relativity (GR) on cosmological scales and the nature of dark energy and dark matter.

Modified gravity theories offer a well-motivated and generic theoretical framework for extending and potentially supplanting the standard $\Lambda$CDM model. These theories, which range from scalar--tensor theories to more exotic formulations such as $f(R,T)$ gravity and non-metricity gravity, propose alterations or extensions to GR that can naturally account for the accelerated expansion of the Universe without resorting to the cosmological constant or {elucidating the dynamics of cosmic structure formation in novel ways}. 
Their appeal lies not only in their ability to address the existing tensions~\cite{Bahamonde:2021gfp,Adi:2020qqf,Marra:2021fvf,Odintsov:2020qzd,Ballardini:2020iws,Wang:2020zfv,Skara:2019usd,Escamilla-Rivera:2019ulu,Cai:2019bdh,Kazantzidis:2019nuh,Kazantzidis:2018rnb,Nunes:2018xbm,Gangopadhyay:2022bsh,Zumalacarregui:2020cjh,Benevento:2022cql,Braglia:2020auw,Farhang:2020sij,Heisenberg:2020xak,Ballesteros:2020sik,FrancoAbellan:2023gec} but also in their potential to enrich our theoretical landscape with new physics that could resolve longstanding puzzles in cosmology~\cite{CANTATA:2021ktz,Langlois:2017dyl,Nojiri:2017ncd,Joyce:2016vqv,Joudaki:2020shz,Koyama:2015vza,Kobayashi:2019hrl,Ishak:2018his,Sakstein:2017xjx,Amendola:2016saw,Crisostomi:2016czh}.

{This} \href{https://www.mdpi.com/journal/universe/special_issues/Gravit}{Special Issue} aims to delve into the heart of these debates, presenting a collection of cutting-edge research that explores the theoretical viability, empirical implications, and observational constraints of modified gravity theories. Through this collective endeavor, we seek to illuminate the pathways toward a deeper understanding of the cosmos, guided by the principle that the resolution of the $\Lambda$CDM tensions could unveil new facets of our Universe and lay the groundwork for a new standard model of cosmology.


\section{Overview of the Published Articles}
\label{sec:overview}

Maria Petronikolou and Emmanuel N. Saridakis's article (contribution 1~\cite{Petronikolou:2023cwu}) delves into scalar--tensor- and bi-scalar--tensor-modified theories of gravity as potential frameworks for alleviating the Hubble tension. By selecting models with a unique shift-symmetric friction term, their work demonstrates how these theories can significantly mitigate the discrepancy between local and early Universe measurements of the Hubble constant, offering new pathways for resolving this longstanding cosmological puzzle.

Ziad Sakr's research (contribution 2~\cite{Sakr:2023bms}) focuses on untangling the $\sigma_8$ discomfort by independently analyzing the matter fluctuation parameter $\sigma_8$ and the growth index $\gamma$. His innovative approach treats $\sigma_8$ as a free parameter, distinct from its traditional derivation, revealing how this separation can lead to more accurate constraints on cosmological parameters and offering a novel perspective on the growth tension within the $\Lambda$CDM~framework.

Joan Solà Peracaula and colleagues (contribution 3~\cite{SolaPeracaula:2023swx}) investigate the Running Vacuum Model (RVM) as a dynamical alternative to the cosmological constant. Their extensive analysis, supported by the latest observational data, showcases the RVM's ability to provide a compelling fit to cosmic phenomena while potentially resolving both the $\sigma_8$ and $H_0$ tensions, marking a significant step towards a dynamic understanding of dark~energy.

Christian Böhmer, Erik Jensko, and Ruth Lazkoz (contribution 4~\cite{Boehmer:2023knj}) employ dynamical systems analysis to explore $f(Q)$ gravity's implications for cosmological evolution. Their study elucidates how modifications in the gravity sector can lead to viable cosmological models that offer new insights into the accelerated expansion of the Universe, challenging the conventional $\Lambda$CDM model with a fresh theoretical perspective.

Mayukh R. Gangopadhyay and colleagues (contribution 5~\cite{Gangopadhyay:2022bsh}) present a scenario of large-scale modification of gravity without invoking extra degrees of freedom. Their model, incorporating interactions between baryonic and dark matter, offers a unified framework to address both the late-time acceleration of the Universe and the Hubble tension, suggesting a seamless integration of dark sector phenomena within modified gravity theories.

Filippo Bouché, Salvatore Capozziello, and Vincenzo Salzano (contribution 6~\cite{Bouche:2023xjw}) tackle cosmological tensions through the lens of non-local gravity. By revisiting the foundations of gravitational interaction, their work highlights how non-local modifications can reconcile discrepancies in $\Lambda$CDM predictions, opening the door to alternative models of dark energy and gravity.

\textls[-5]{Celia Escamilla-Rivera and Rubén Torres Castillejos (contribution 7~\cite{Escamilla-Rivera:2022mkc}) explore the Hubble tension using supermassive black hole shadows data. Their innovative approach leverages high-resolution astrophysical observations to infer $H_0$, demonstrating the potential of black hole shadows as novel probes in cosmology and offering a fresh avenue for resolving observational tensions.}

Denitsa Staicova's contribution (contribution 8~\cite{Staicova:2022zuh}) examines dynamical dark energy models in light of combined $H_0 \cdot r_d$ measurements. By analyzing the multiplication of the Hubble parameter and sound horizon scale as a single parameter, Staicova provides new insights into dark energy dynamics, suggesting that dynamical models could offer a resolution to the Hubble tension.

Savvas Nesseris (contribution 9~\cite{Nesseris:2022hhc}) reviews the effective fluid approach as a versatile framework for representing a wide array of modified gravity models. By treating modified gravity effects as contributions from an effective dark energy fluid, Nesseris provides a unified analysis tool for comparing theoretical predictions with observational data. This approach not only facilitates the examination of modified gravity's role in cosmic acceleration but also provides a systematic method for confronting these models with the growth rate of structure and the expansion history of the Universe, offering a bridge between theory and observation.

V. K. Oikonomou, Pyotr Tsyba, and Olga Razina (contribution 10~\cite{Oikonomou:2022tjm}) offer a novel perspective by exploring how Earth's geological and climatological history, alongside the shadows of galactic black holes, might reveal insights into our Universe's evolution. Their interdisciplinary approach highlights potential signatures of past pressure singularities and their implications for the early Universe's dynamics. This intriguing exploration underscores the untapped potential of combining astrophysical, geological, and climatological data to inform and constrain cosmological models, opening up new avenues for understanding the Universe's past and the nature of gravity.

Sunny Vagnozzi (contribution 11~\cite{Vagnozzi:2023nrq}) presents a compelling argument that early-time physics modifications alone may be insufficient to resolve the Hubble tension. By synthesizing evidence across seven hints derived from cosmological observations and theoretical considerations, Vagnozzi's work critically assesses the landscape of proposed solutions to the Hubble tension. This contribution emphasizes the necessity for a comprehensive approach that accounts for both early- and late-time Universe physics, challenging the cosmological community to broaden the scope of theoretical explorations in search of a more complete resolution to one of modern cosmology's most pressing puzzles.

Together, these eleven contributions exemplify this Special Issue's commitment to exploring the frontiers of modified gravity theories as viable alternatives or extensions to the $\Lambda$CDM model. Each article advances the dialogue within the cosmological community, offering fresh insights, innovative methodologies, and compelling arguments that collectively enrich our understanding of the Universe's fundamental nature and its governing~laws.

\section{Conclusions}
\label{sec:conclusion}

The Special Issue ``Modified Gravity Approaches to the Tensions of $\Lambda$CDM'' represents a significant stride toward addressing some of the most perplexing challenges in modern cosmology. Through a diverse collection of contributions, this endeavor has not only contributed to the illumination of the intricacies of the tensions within the $\Lambda$CDM model, but has also underscored the potential of modified gravity theories to pave the way for groundbreaking discoveries. Each article, in its unique capacity, has contributed to a deeper understanding of the Universe's fundamental laws, showcasing the richness and vitality of theoretical innovation in the face of empirical puzzles.

The collective progress made through this Special Issue is a testament to the importance of continued exploration and open-mindedness in the scientific quest to understand the cosmos. The detailed examinations of modified gravity theories presented here underscore the necessity of extending our theoretical frameworks beyond the confines of general relativity and the standard cosmological model. These theories offer not just solutions to specific observational tensions, but also a broader perspective on gravity and its role in the evolution of the Universe.

Moreover, this Special Issue stands as a potential milestone in the discovery of new physics. The tensions within the $\Lambda$CDM model may indeed signal the limitations of our current understanding, pointing toward an underlying reality that is more complex and nuanced than previously thought. In this context, modified gravity theories emerge not merely as alternatives, but as harbingers of a new standard model of cosmology. They represent the most fundamental approach to addressing the cosmological tensions, weaving together the empirical anomalies with theoretical insights to sketch a more accurate and comprehensive picture of the Universe.

The journey toward resolving the enduring tensions within the $\Lambda$CDM model and unveiling the new physics that will undoubtedly reshape our cosmological paradigm is far from over. However, the contributions of this Special Issue mark critical waypoints on this journey. They invite the scientific community to reconsider the foundations of cosmology, to embrace the uncertainties and anomalies, not as mere nuisances, but as beacons guiding us toward a more profound understanding of the Universe.

As we continue to probe the depths of the cosmos with ever more sophisticated tools and technologies, the insights gleaned from modified gravity theories will undoubtedly play a special role. They offer a promising path forward, one that harmonizes the elegance of theoretical physics with the empirical realities of the Universe we strive to understand. In this endeavor, the Special Issue ``Modified Gravity Approaches to the Tensions of $\Lambda$CDM'' stands as both a milestone and a beacon, illuminating the path towards the next standard model of cosmology and the new physics that will underpin it.
\vspace{6pt}

\conflictsofinterest{The authors declare no conflicts of interest.}


\begin{adjustwidth}{-\extralength}{0cm}
\reftitle{References}

\PublishersNote{}
\end{adjustwidth}


\end{document}